\pdfoutput=1
\documentclass[english,showkeys]{revtex4-1}
\usepackage[T1]{fontenc}
\usepackage[latin9]{inputenc}
\usepackage{xcolor}
\usepackage{pdfcolmk}
\usepackage{babel}
\usepackage{units}
\usepackage{amsbsy}
\usepackage{amstext}
\usepackage{graphicx}
\PassOptionsToPackage{normalem}{ulem}
\usepackage{ulem}
\usepackage[unicode=true,pdfusetitle,
 bookmarks=true,bookmarksnumbered=false,bookmarksopen=false,
 breaklinks=false,pdfborder={0 0 0},pdfborderstyle={},backref=false,colorlinks=true]
 {hyperref}

\makeatletter

\newcommand{\noun}[1]{\textsc{#1}}
\providecolor{lyxadded}{rgb}{0,0,1}
\providecolor{lyxdeleted}{rgb}{1,0,0}
\DeclareRobustCommand{\lyxadded}[3]{{\texorpdfstring{\color{lyxadded}{}}{}#3}}

\@ifundefined{showcaptionsetup}{}{%
 \PassOptionsToPackage{caption=false}{subfig}}
\usepackage{subfig}
\makeatother

\begin{document}

\title{\noun{Vegas}: Software package for the atomistic simulation of magnetic
materials}

\author{J.D. Alzate-Cardona}
\email{Corresponding author. Tel: +57 6 8879300 Ext. 55631. E-mail: jdalzatec@unal.edu.co (Juan David Alzate-Cardona)}

\author{D. Sabogal-Suárez}

\author{O.D. Arbeláez-Echeverri}

\author{E. Restrepo-Parra}

\affiliation{Departamento de Física y Química~\\
Universidad Nacional de Colombia,~\\
Sede Manizales, A.A. 127 Manizales, Colombia.}
\begin{abstract}
We present an open-source software package, \noun{Vegas}, for the
atomistic simulation of magnetic materials. Using the classical Heisenberg
model and the Monte Carlo Metropolis algorithm, \noun{Vegas} provides
the required tools to simulate and analyze magnetic phenomena of a
great variety of systems. \noun{Vegas} stores the history of the simulation,
i.e. the magnetization and energy of the system at every time step,
allowing to analyze static and dynamic magnetic phenomena from results
obtained in a single simulation. Also, standardized input and output
file formats are employed to facilitate the simulation process and
the exchange and archiving of data. We include results from simulations
performed using \noun{Vegas}, showing its applicability to study different
magnetic phenomena.
\end{abstract}

\maketitle

\section{Introduction}

Magnetic materials are widely used in a diverse range of applications
in modern society. In recent years, magnetic nanostructures have attracted
much attention because of their promising properties that are no observed
in larger structures and the miniaturization demand of magnetic devices.
Properties like exchange bias and superparamagnetic behavior at room
temperature make magnetic nanostructures ideal for magnetic recording
media \citep{Nordblad2015} and biomedical applications \citep{Neuberger2005},
respectively. 

Spin models have played an important role in understanding the magnetic
behavior of magnetic materials. Specially, classical spin models bridge
the gap between a full electronic description of a magnetic material
and conventional micromagnetism \citep{Nowak2007}. Classical spins
can be represented as $n$ dimensional vectors, such as Ising ($n=1$),
XY ($n=2$) and Heisenberg ($n=3$) spins. Because of the difficulty
to solve by analytical approaches the partition function of systems
represented by classical spin models, numerical simulations employing
the Monte Carlo method are usually used to estimate the thermodynamic
quantities of the systems. Magnetic properties of different systems,
such as thin films \citep{Sabogal-Suarez2015}, simple and core-shell
nanoparticles \citep{Evans2011,Alzate-Cardona2017b}, mixed spin systems
\citep{Alzate-Cardona2017}, torus nanorings \citep{Alzate-Cardona2017a},
nanotubes \citep{Salazar-Enriquez2013}, bit-patterned media \citep{Arbelaez-Echeverri2016}
and bulk materials \citep{Restrepo-Parra2011,Agudelo-Giraldo2015}
have been studied by Monte Carlo simulations. These atomistic simulations
allow to take into account changes in the magnetization that occur
at atomic scale and the finite size effects which are considerable
in magnetic nanomaterials \citep{Evans2014}.

Because of the complexity of the methods and algorithms used to perform
atomistic simulation of magnetic materials, research in the area is
usually restricted to experts with advanced knowledge in computer
programming, where simulations are mostly based on codes developed
by the researchers themselves. In view of that, different open source
software packages, such as \noun{Vampire \citep{Evans2014}} and ALPS
\citep{Alet2005}, have been developed with the aim to make available
these kind of simulations to the non-expert software developer. Furthermore,
the development new software is important to take into account improvements
made to the current methods and implementations. An increase in the
efficiency of the simulation and analysis processes can be achieved
by the implementation of novel algorithms and the enhancement of the
structure employed to describe the simulation input and output data.
Besides, just part of the data is usually collected during a simulation,
requiring several simulations to study distinct magnetic behavior
of a given system. Taking these aspects into account, we present an
open source software package, \noun{Vegas}, for the static and dynamic
atomistic simulation of magnetic materials.

\section{VEGAS software package}

The aim of \noun{Vegas} is to provide the required tools to build,
simulate and analyze a great variety of magnetic systems with structural
and magnetic characteristics that can be set with high flexibility.
The main features of \noun{Vegas} are the following:
\begin{itemize}
\item The history of the simulation is stored, i.e. the magnetization and
energy of the system are stored at every Monte Carlo Step (MCS) per
temperature and magnetic field step. This allows to study the dynamic
behavior of the system in addition to its static behavior from results
obtained in a single simulation. In this way, it is possible to estimate,
among other things, equilibrium and non-equilibrium correlation times
which can be used to enhance the quality of the statistics of the
thermal averages. 
\item High flexibility is provided for the building and simulation of a
great variety of magnetic systems. It is possible to simulate systems
with different structures and magnetic phases, where the statistics
of each magnetic phase can be tracked independently.
\item Several spin update policies, including a highly-efficient adaptive
policy\lyxadded{Juan David Alzate Cardona}{Wed Jan 24 19:34:19 2018}{},
are implemented in VEGAS to simulate accurately the magnetic behavior
of different magnetic systems. It is possible to simulate a material
with various magnetic phases employing a different update policy for
each magnetic phase, e.g. a core/shell nanoparticle where the core
is updated using a random move and the shell is updated using a spin-flip
move. Also, Ising and mixed spin Ising systems can be modeled using
the spin-flip and qIsing update policies, respectively.
\item The direction of every spin moment at the last MCS is stored per temperature
and magnetic field step. Hence, it is possible to visualize the temperature/field
evolution of the spin moments direction, which can contribute to a
better understanding of the magnetic phenomena.
\item An estimation of the simulation time is given during the simulation.
\end{itemize}

\section{Model and method}

\subsection{Classical spin Hamiltonian}

The magnetic systems simulated using \noun{Vegas} are described by
the classical Heisenberg model. Interactions at the atomic scale between
spin moments in a magnetic material are modeled by the Heisenberg
spin Hamiltonian, which includes the exchange, anisotropy and applied
field interactions. The Heisenberg spin Hamiltonian is given by

\begin{equation}
\mathcal{H}=-\sum_{i\neq j}J_{ij}\text{\textbf{S}}_{i}\cdot\text{\textbf{S}}_{j}-B\sum_{i}\left(\text{\textbf{S}}_{i}\cdot\boldsymbol{n}_{i}\right)-\mathcal{H}_{ani}\label{eq:1}
\end{equation}

where $J_{ij}$ is the exchange interaction constant between sites
$i$ and $j$, $B$ and $\boldsymbol{n}_{i}$ are the magnetic field
intensity and direction of site $i$, respectively, $\text{\textbf{S}}_{i}$
and $\text{\textbf{S}}_{j}$ are the spin moment directions of sites
$i$ and $j$, respectively, and $\mathcal{H}_{ani}$ is the anisotropy
term. The anisotropy term can take the form of uniaxial anisotropy

\begin{equation}
\mathcal{H}_{ani}^{uni}=\sum_{i}k_{i}^{uni}\left(\text{\textbf{S}}_{i}\cdot\boldsymbol{e}_{i}\right)^{2}
\end{equation}
where $k_{i}^{uni}$ and $\boldsymbol{e}_{i}$ are the uniaxial anisotropy
constant and direction of site $i$, respectively, or cubic anisotropy
\citep{Skomski2008}

\begin{equation}
\mathcal{H}_{ani}^{cub}=\sum_{i}k_{i}^{cub}\left(S_{x}^{2}S_{y}^{2}+S_{y}^{2}S_{z}^{2}+S_{x}^{2}S_{z}^{2}\right)_{i}
\end{equation}
where $k_{i}^{cub}$ is the cubic anisotropy constant, and $S_{x}$,
$S_{y}$ and $S_{z}$ are the $x$, $y$ and $z$ components of the
spin moment $\mathbf{S}$, respectively.

\subsection{Monte Carlo Metropolis algorithm}

The time evolution of a classical spin magnetic system can be given
by the Monte Carlo Metropolis algorithm, where new configurations
of a system are generated from a previous state using a transition
probability. The algorithm is developed as follows: an initial state
is chosen for the magnetic system. Then, one single spin moment is
randomly chosen and its direction is changed to a new trial direction
according to an update policy (trial move). The energy of the system
before and after the trial move is then evaluated according to Eq.
(\ref{eq:1}) and the energy change $\left(\Delta E\right)$ is calculated.
Finally, the trial move is accepted with a probability given by

\begin{equation}
p=\exp\left(-\frac{\Delta E}{k_{B}T}\right)
\end{equation}
where $k_{B}$ is the Boltzmann constant and $T$ is the absolute
temperature of the system. Repeating this process once per spin moment
is called a MCS. The update policy has influence on the efficiency
of the algorithm and its physical interpretation \citep{Nowak2007}.
Common update policies, including the spin-flip, random, small step
and Gaussian moves \citep{Hinzke1999}, are implemented in \noun{Vegas}.
Furthermore, an adaptive move which keeps the acceptance rate of the
new states near $50\%$, enhancing the efficiency of the Monte Carlo
Metropolis algorithm is implemented.

For a comprehensive review of the atomistic spin model and the integration
methods employed in the atomistic simulation of magnetic materials
see \citep{Evans2014}.

\subsection{File Formats}

\begin{figure}[h]
\centering{}\includegraphics[width=0.3\columnwidth]{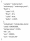}\caption{Example of a JSON input file defining the simulation parameters. \label{fig:JSON}}
\end{figure}

In order to facilitate the exchange and archiving of data, \noun{Vegas}
uses the standardized file formats, \noun{Json} and \noun{Hdf5,} for
the input and output files of the simulations, respectively. \noun{Json}
uses universal data structures through which the simulation parameters
can be defined using a name/value pair structure. An example of a
\noun{Json} input file is shown in Figure \ref{fig:JSON}. In this
file, the simulation parameters, including the temperature, magnetic
field range and step, number of MCS and random number seed, for a
hysteresis loop simulation are defined. Also, the material, anisotropy
and initial state (if required) files, and the results output file
are given. The simulation parameters are normalized such that $k_{B}=1$.
However, if the user intends to use real units or other kind of normalization,
it is necessary to use coherent units for the exchange interaction,
anisotropy constant, magnetic field intensity and Boltzmann constant.
On the other hand, \noun{Hdf5} is a data model and file format for
storing and managing extremely large data, which is useful for storing
and analyzing the simulation results in \noun{Vegas}, specially the
history of the simulation. Other input data, including the anisotropy
and the material files, are given in plain text format. For these
files, we designed simple data schemes to specify the anisotropy type
and the magnetic and structural parameters of the material.

\subsection{\noun{Vegas} tools}

Based on different \noun{python} and \noun{C++} libraries, \noun{Vegas}
provides tools to build, simulate and analyze magnetic systems. Tutorials
and a detailed description of the use of \noun{Vegas} can be found
at its web page \emph{\href{https://pcm-ca.github.io/vegas/}{https://pcm-ca.github.io/vegas/}}.

\subsubsection{Material building}

\noun{Vegas} has a \noun{python}-based tool, \noun{Vegas lattice},
for building materials with different shapes, structures and magnetic
properties. \noun{Vegas lattice} provides the essential routines to
generate regular graph lattices in linear time, as well as some cuts
of those lattices for nanoparticles and randomly depleted lattices.
Detailed documentation of \noun{Vegas lattice} can be found at its
GitHub repository \citep{Arbelaez-Echeverri}. 

It is also possible to build the material manually according to the
material file format. The material file, which uses a plain text format,
is essentially composed by four parts. The first part indicates the
number of ions, links and ion types in the material. The second part
is composed by identifiers for the different types of ions. The third
part mainly corresponds to the geometric construction of the material,
indicating the index, spatial coordinates, spin norm, spatial coordinates
of the magnetic field axis, type and update policy of the ion. Finally,
the fourth part indicates the exchange interactions between the ions.

\subsubsection{Simulation}

The simulation model and methods are implemented in \noun{Vegas} using
\noun{C++} libraries. In order to initialize the simulation of a given
magnetic material,\noun{ Vegas }reads the system and simulation parameters
from the \noun{Json} input file. Then, the simulation is carried out
by routines implemented according to the classical Heisenberg model
and the Monte Carlo Metropolis algorithm, while the simulation results
are stored in the \noun{Hdf5} output file. During the simulation,
an estimate of the simulation time is given.

\subsubsection{Data analysis}

\noun{Vegas} has implemented multiple analyzers for the analysis and
visualization of different magnetic phenomena. All the analyzers are
written in \noun{python} using the \noun{h5py} library. The analyzers
are developed based on the data structure of the simulation results,
which are stored in \noun{Hdf5} format. Different variables are employed
to store the simulation result data, which include the temperature,
magnetic field, ion positions, final states, energy and magnetization.
Furthermore, the Boltzmann constant and the number of MCS are stored
as attributes. 

If the user requires a more specific or detailed analysis, it is possible
to implement simple scripts to analyze the simulation results. Moreover,
the HDF Group provides a visual tool, \noun{HDFView} \citep{Hdfview},
to extract and analyze data from \noun{Hdf5} files.

\section{Applications}

Static and dynamic magnetic properties, such as hysteresis loops,
critical temperature, compensation behavior, reversal mechanisms,
magnetic recording, critical exponents and correlation times, can
be determined using \noun{Vegas}. In this Section, we present some
examples of the applicability of \noun{Vegas} to study different materials
and magnetic phenomena.

\subsection{Materials with different shapes and structures}

\begin{figure}[h]
\begin{centering}
\subfloat[]{\includegraphics[width=0.3\columnwidth]{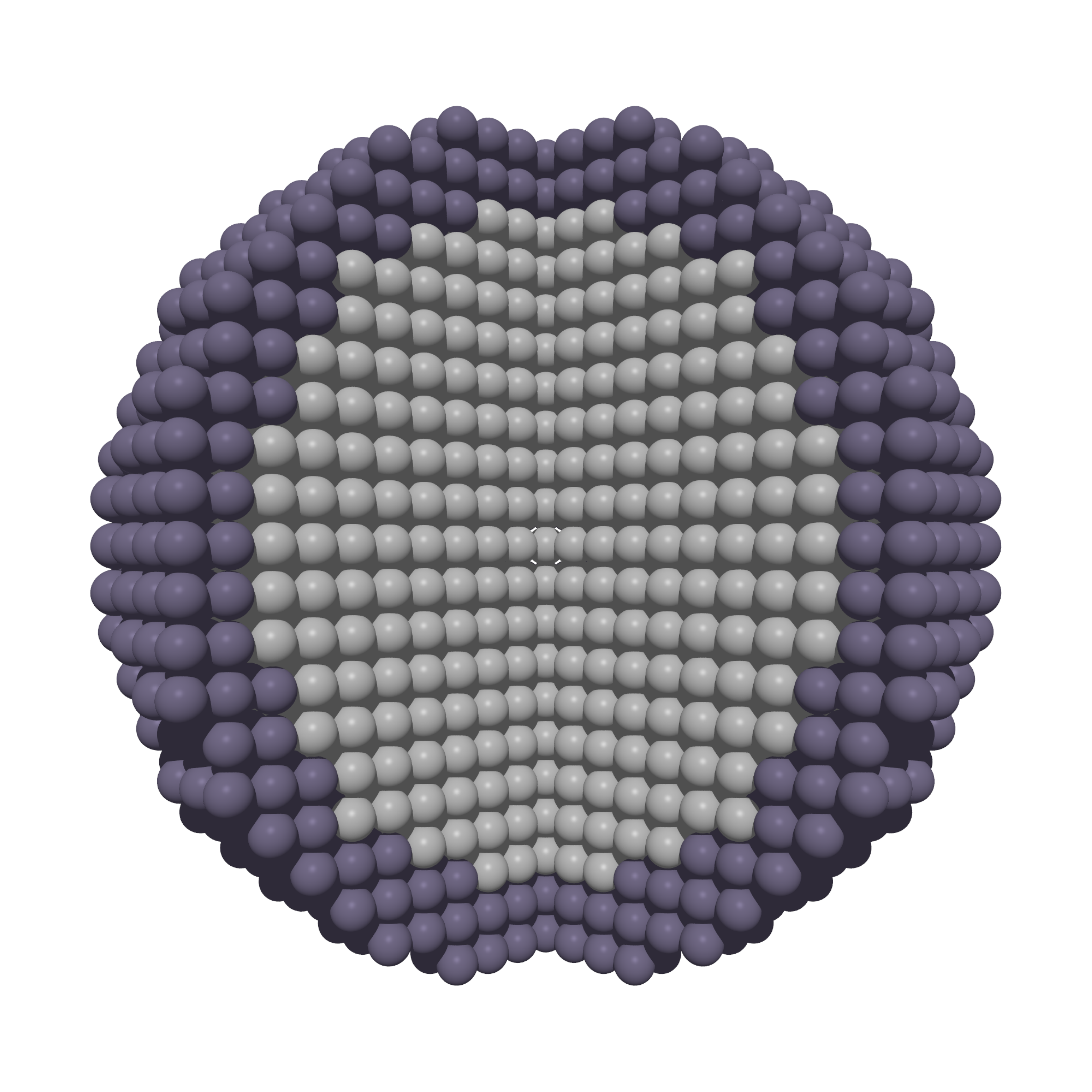}

}\subfloat[]{\includegraphics[width=0.3\columnwidth]{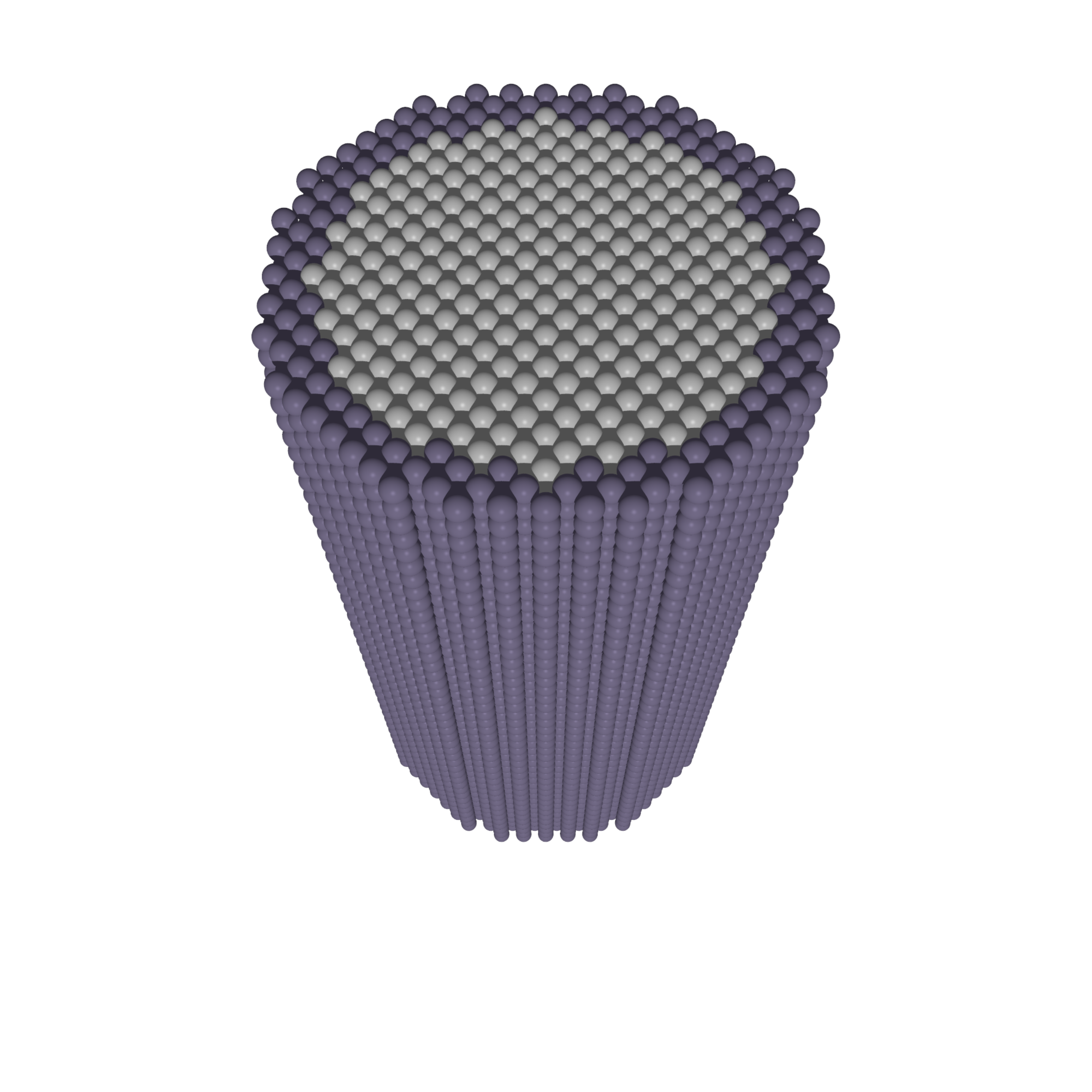}

}\subfloat[]{\includegraphics[width=0.3\columnwidth]{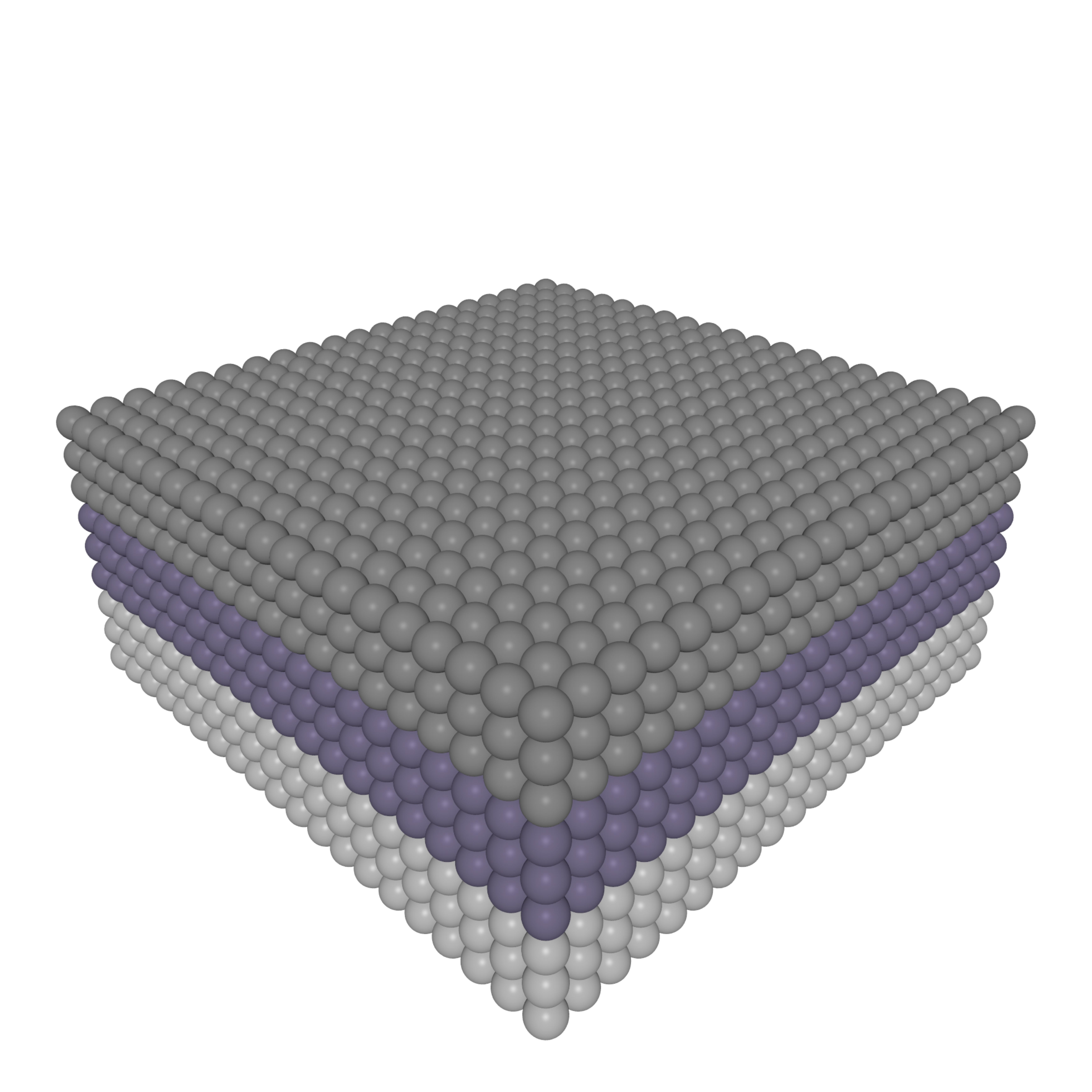}

}
\par\end{centering}
\caption{Samples of a (a) spherical core/shell nanoparticle, (b) core/shell
nanowire and (c) multilayer. \label{fig:materials}}

\end{figure}

Materials with simple and complex shapes, such as core/shell nanoparticles,
nanotubes, nanowires, nanorings, bit patterned media, thin films,
multilayers and bulk systems, and crystal structures, such as simple,
body-centered and face-centered cubic, can be simulated in \noun{Vegas.}
Samples of some materials are shown in Figure \ref{fig:materials}.

\subsection{Thermal averages}

\begin{figure}[h]
\begin{centering}
\subfloat[\label{fig:irona}]{\includegraphics[width=0.4\columnwidth]{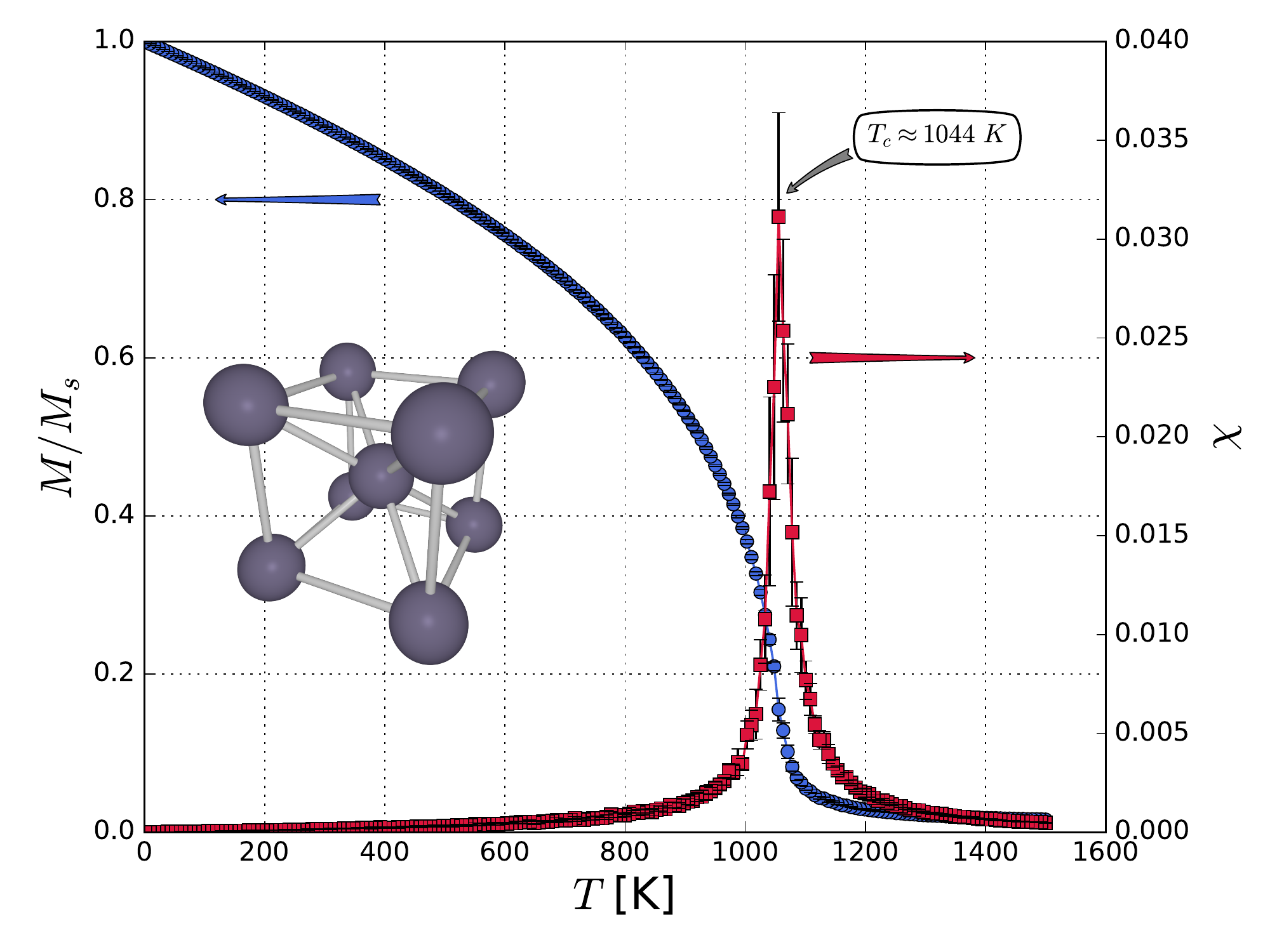}

}\subfloat[]{\includegraphics[width=0.4\columnwidth]{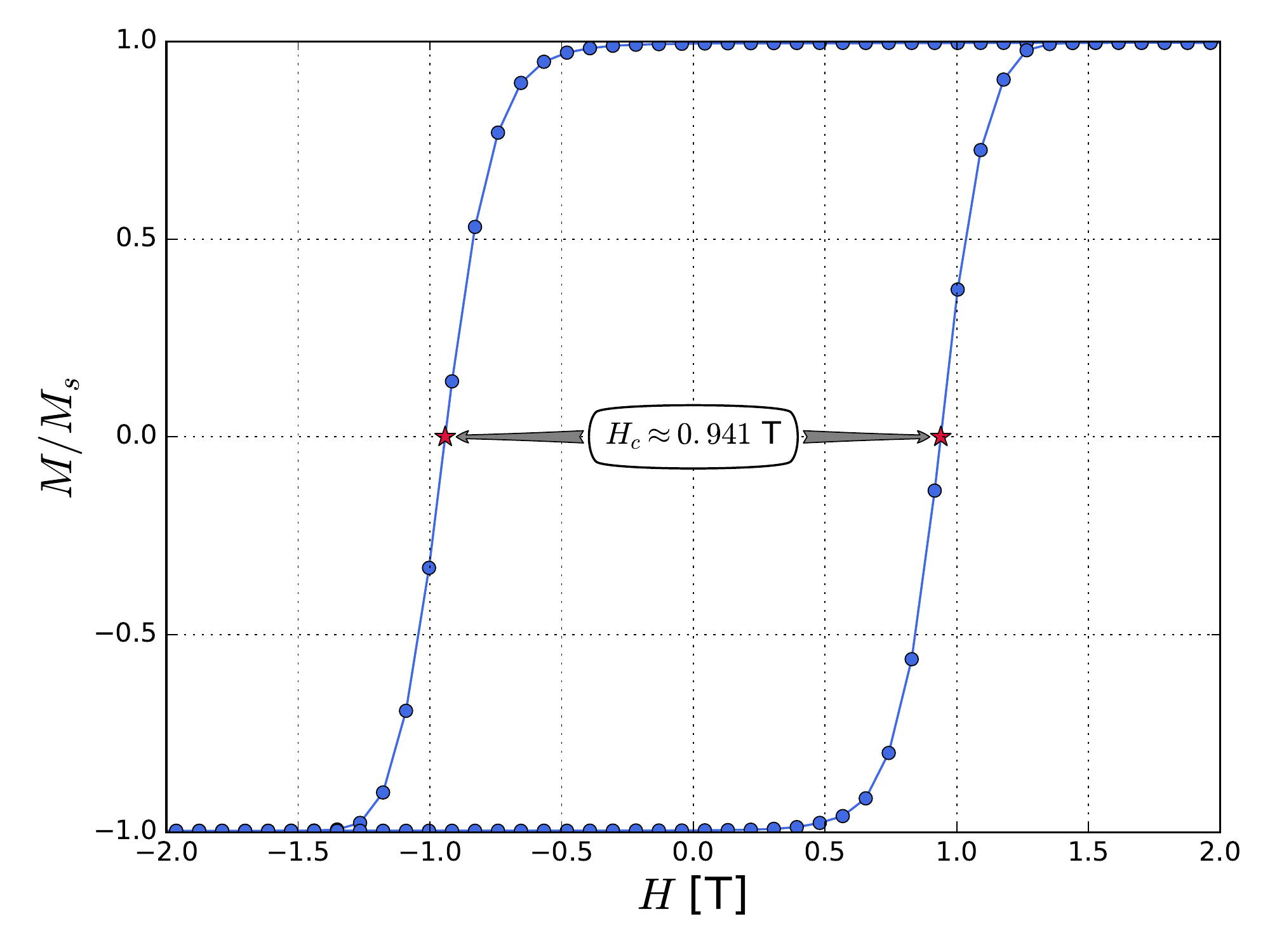}

}
\par\end{centering}
\caption{(a) Thermal dependence of the magnetization and susceptibility, and
(b) hysteresis loop at $10\:K$ of an iron bulk material. Inset in
Figure \ref{fig:irona} shows the BCC crystal structure of iron. \label{fig:iron}}

\end{figure}

Thermal averages can be easily extracted and visualized from the simulation
results output file. Figure \ref{fig:iron} shows the thermal dependence
of the magnetization ($M$) and the susceptibility ($\chi$), and
a hysteresis loop of a bulk material simulated using the structure
and magnetic parameters of iron, namely, $J=44.01\:\text{\ensuremath{\nicefrac{meV}{link}}}$
and $k=3.53\times10^{-3}\:\nicefrac{\text{meV}}{\text{atom}}$ \citep{Evans2014}.
From these results, important magnetic properties of the material,
such as the critical temperature ($T_{c}$) and the coercivity $H_{c}$,
can be determined.

\subsection{Dynamic behavior}

\begin{figure}[h]
\begin{centering}
\includegraphics[width=0.4\columnwidth]{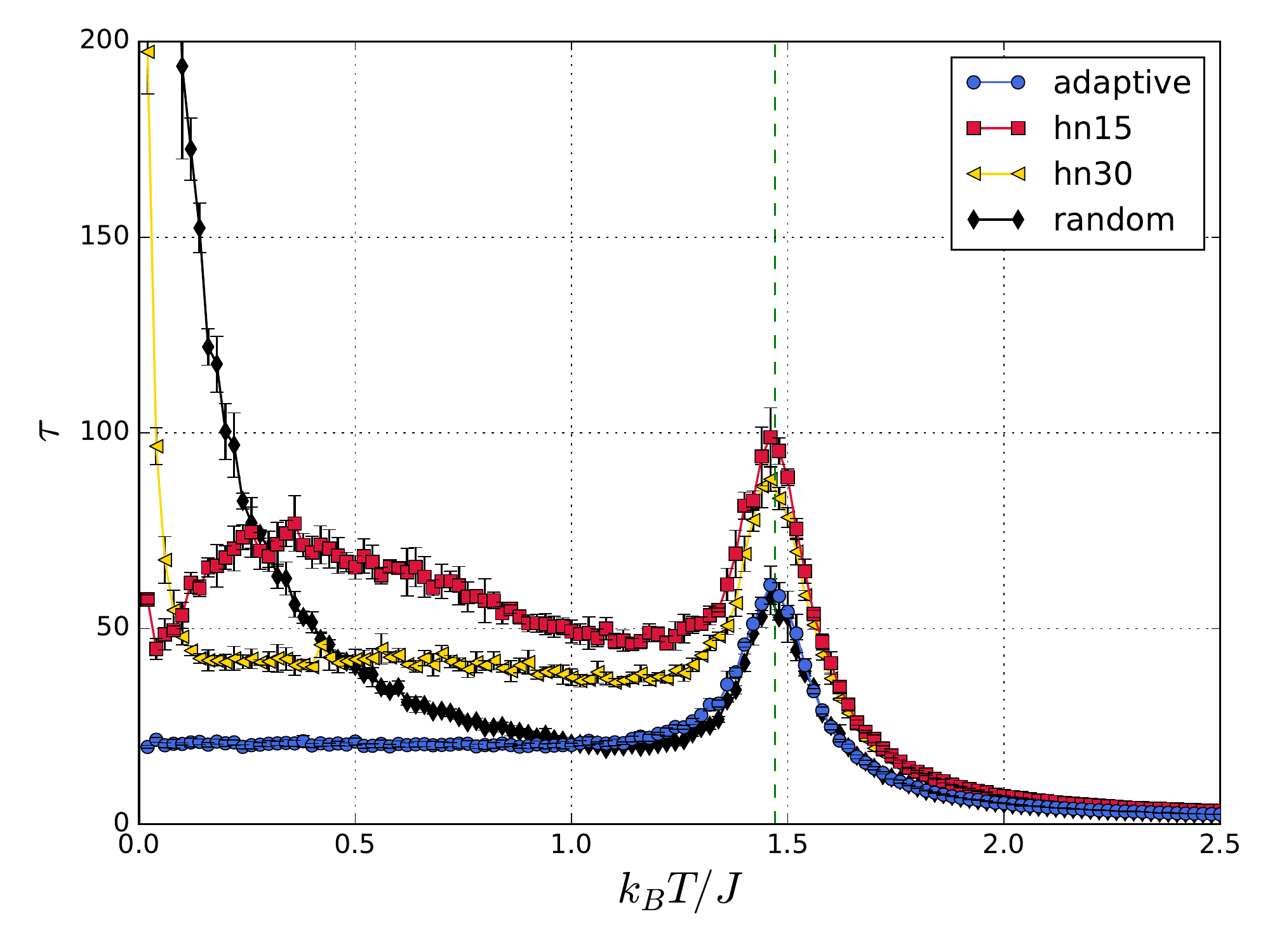}
\par\end{centering}
\caption{Equilibrium correlation times for a bulk material using adaptive,
Hinzke-Nowak combinational \citep{Hinzke1999} (for a small-step opening
angle of $15$ and $30\text{º}$) and random update policies.\label{fig:correlationtimes}}
\end{figure}

Besides the static magnetic properties, the dynamic properties of
a magnetic system can be extracted from a single simulation \citep{Landau2009}.
Figure \ref{fig:correlationtimes} shows the equilibrium correlation
times ($\tau$) for a ferromagnetic bulk material with generic parameters.
The correlation times were computed for some of the different spin
update policies implemented in \noun{Vegas}. Overall, the adaptive
policy produces the lowest correlation times at all temperatures,
indicating that it is a very efficient update policy for this kind
of system.

\subsection{Critical exponents}

\begin{figure}[h]
\begin{centering}
\subfloat[\label{fig:critical_exponents_a}]{\includegraphics[width=0.4\columnwidth]{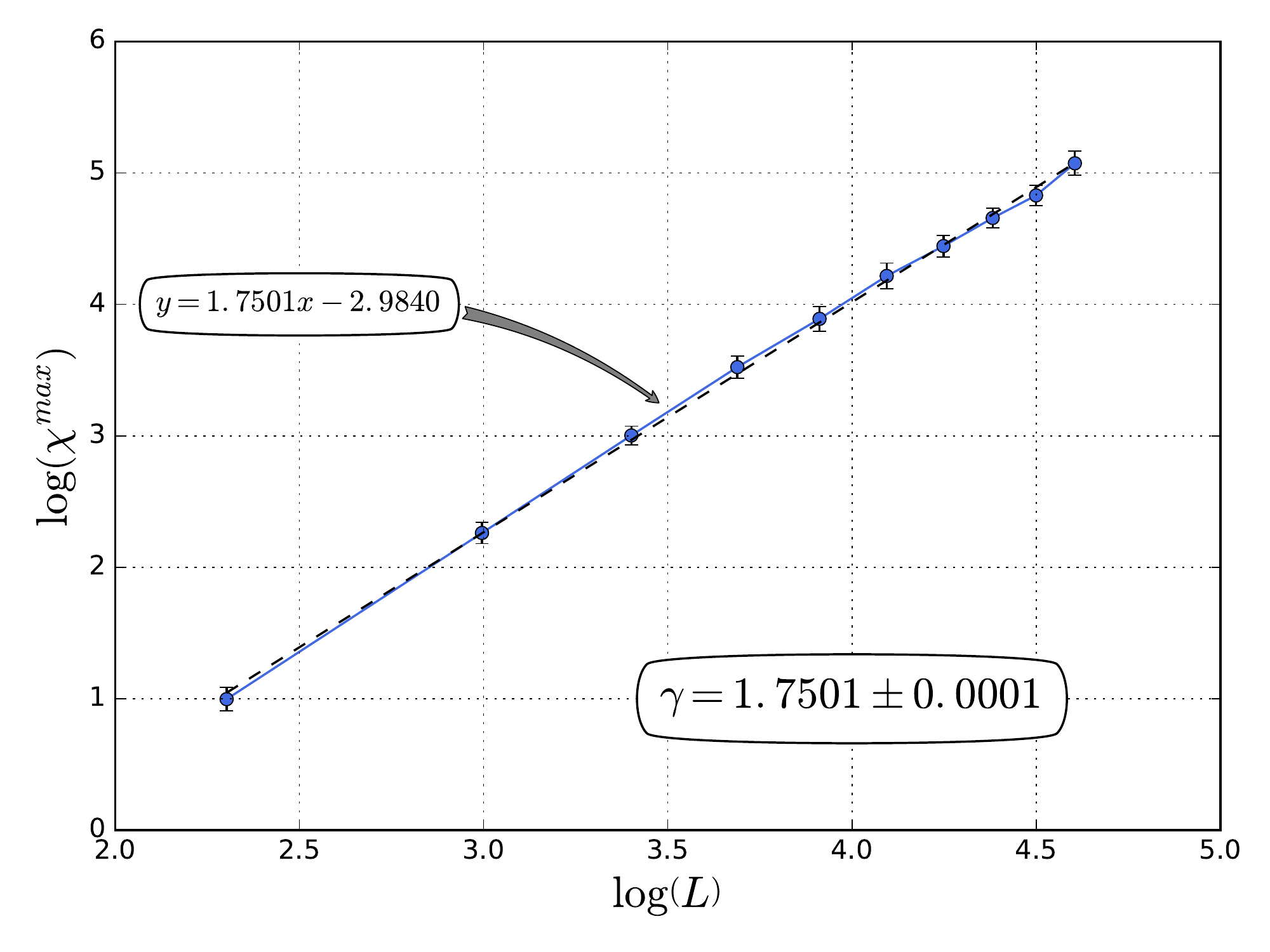}

}\subfloat[\label{fig:critical_exponents_b}]{\includegraphics[width=0.4\columnwidth]{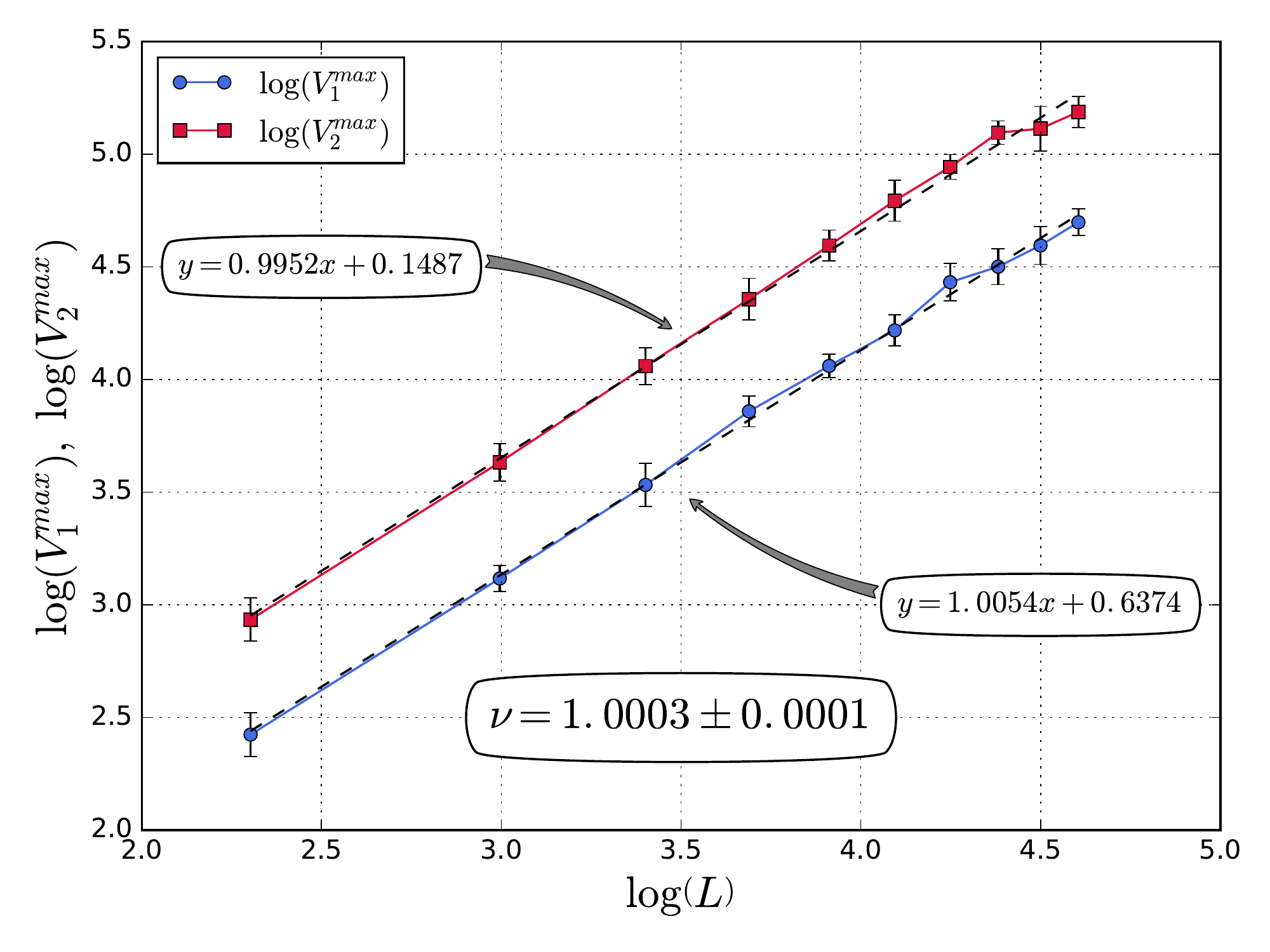}

}
\par\end{centering}
\caption{Log-log plots of (a) the maxima of the susceptibility, and (b) the
first and second order cumulants of the magnetization as a function
of system size for a 2D Ising system. \label{fig:critical_exponents}}
\end{figure}

Phase transitions can be characterized by simulating the critical
behavior of the magnetic systems. Using the finite size scaling method,
it is possible to extract values for critical exponents by observing
the variation of the thermal averages with the system size \citep{Landau2009,Newman1999}.
Figure \ref{fig:critical_exponents} shows log-log plots of the maxima
of the susceptibility ($\chi^{max}$), and the first ($V_{1}^{max}$)
and second order ($V_{2}^{max}$) cumulants of the magnetization as
a function of the system size $(L)$ for a ferromagnetic 2D Ising
system with generic parameters. The critical exponents $\gamma$ and
$\nu$ are obtained from the slope of the plots in Figures \ref{fig:critical_exponents_a}
and \ref{fig:critical_exponents_b}, respectively.

\subsection{Magnetization switching processes}

\begin{figure}[h]
\begin{centering}
\subfloat[\label{fig:torus_a}]{\includegraphics[width=0.3\columnwidth]{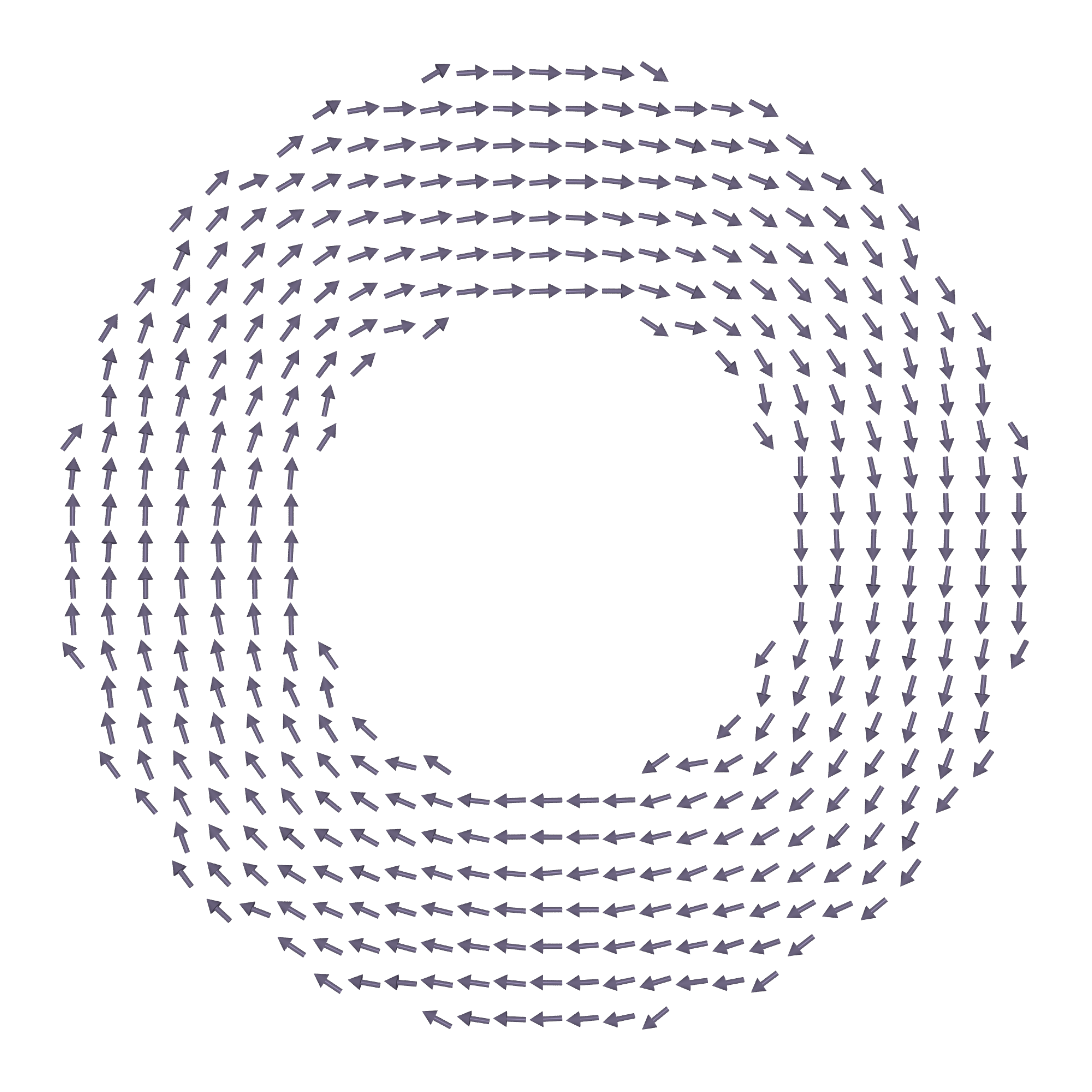}

}\subfloat[\label{fig:torus_b}]{\includegraphics[width=0.3\columnwidth]{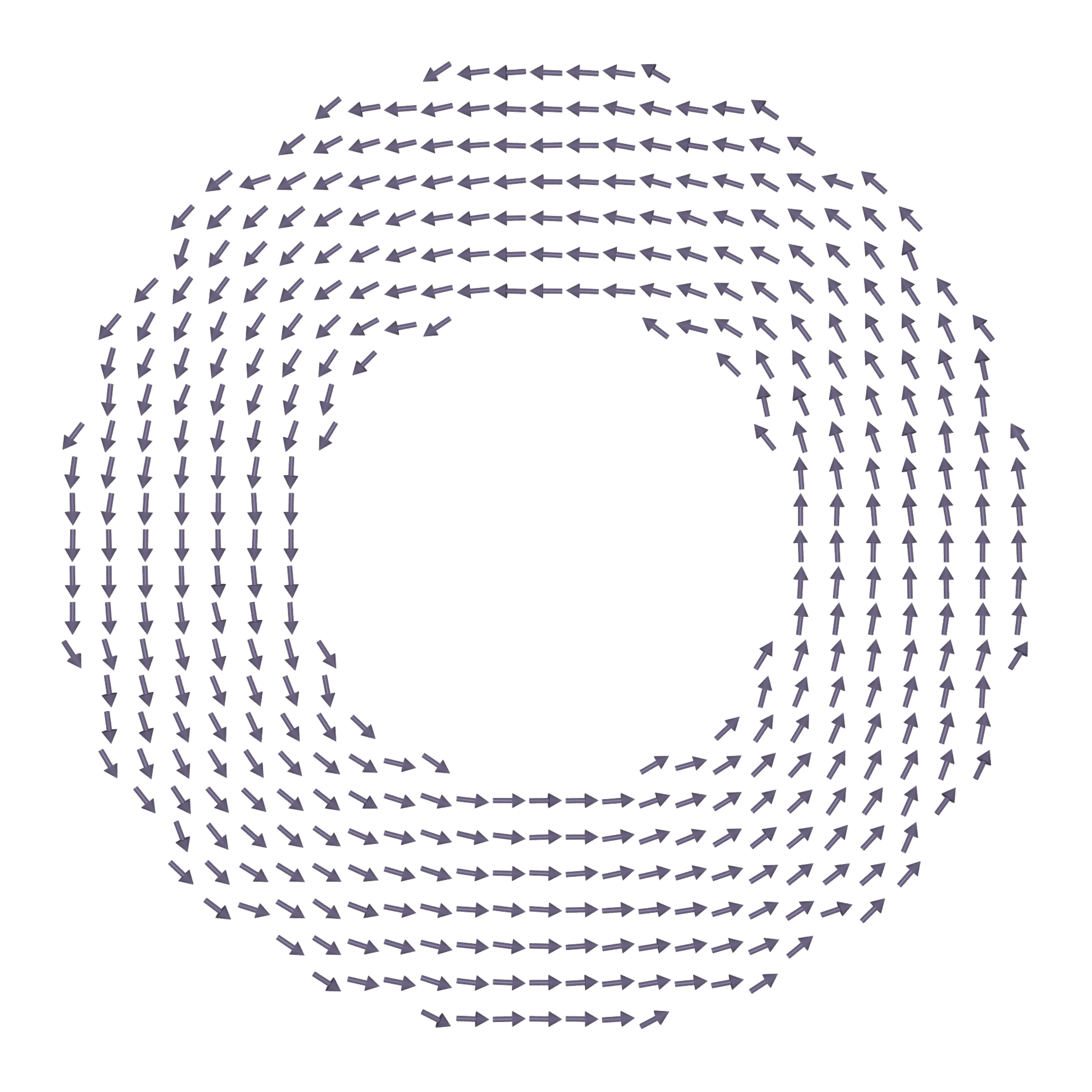}

}
\par\end{centering}
\begin{centering}
\subfloat[\label{fig:torus_c}]{\includegraphics[width=0.4\columnwidth]{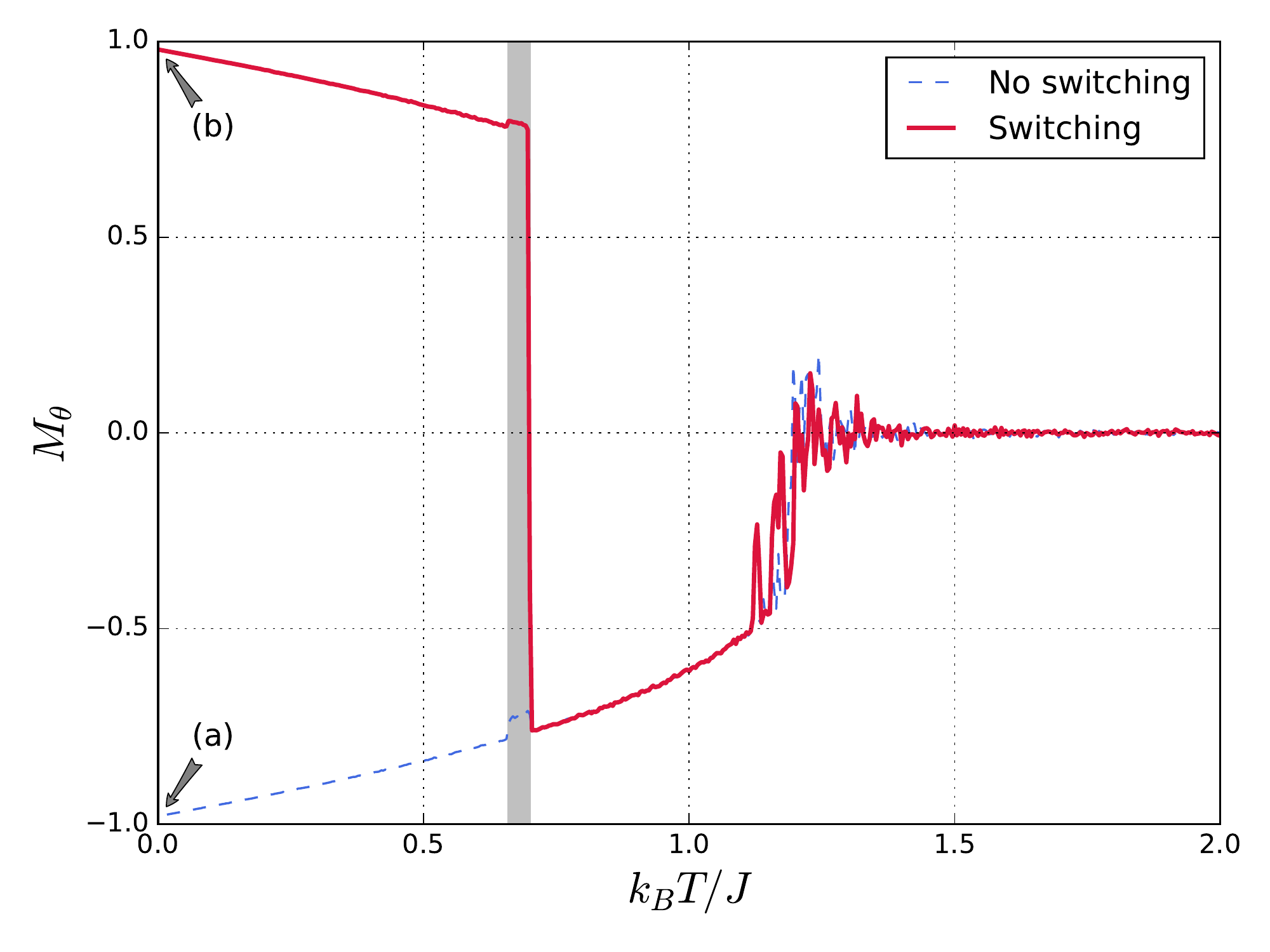}

}
\par\end{centering}
\caption{(a) Vortex and (b) reverse vortex magnetization states in a torus
nanoring, and (c) temperature dependence of the magnetization in the
$\hat{\theta}$ direction of a torus nanoring cooled down from a high
temperature.}
\end{figure}

Simulation of magnetization switching processes are of great importance
for the development of recording devices based on magnetic nanoparticles.
Figures \ref{fig:torus_a} and \ref{fig:torus_b} show cross-sectional
views of magnetization states in a ferromagnetic torus nanoring. Two
magnetization states, vortex (see Figure \ref{fig:torus_a}) and reverse
vortex (see Figure \ref{fig:torus_b}) states, can be formed in the
torus nanoring and conveniently employed to represent the bits 0 and
1. Figure \ref{fig:torus_c} shows the temperature dependence of the
magnetization in the $\hat{\theta}$ direction ($M_{\theta}$) of
two independent simulations where a torus nanoring with generic parameters
was cooled down from a high temperature. During the cooling down process
and in a short temperature range (shaded region), an external circular
magnetic field was applied. In one of the simulations, the external
magnetic field switches the magnetization of the torus nanoring producing
a reverse vortex state (solid line), while in the other simulation,
it does not switch the magnetization producing a vortex state (dashed
line). This kind of behavior indicates that there is a switching probability
which depends on the temperature range at which the external magnetic
field is applied \citep{Alzate-Cardona2017a}.

\subsection{Histogram and multiple histogram methods}

\begin{figure}[h]
\includegraphics[width=0.4\columnwidth]{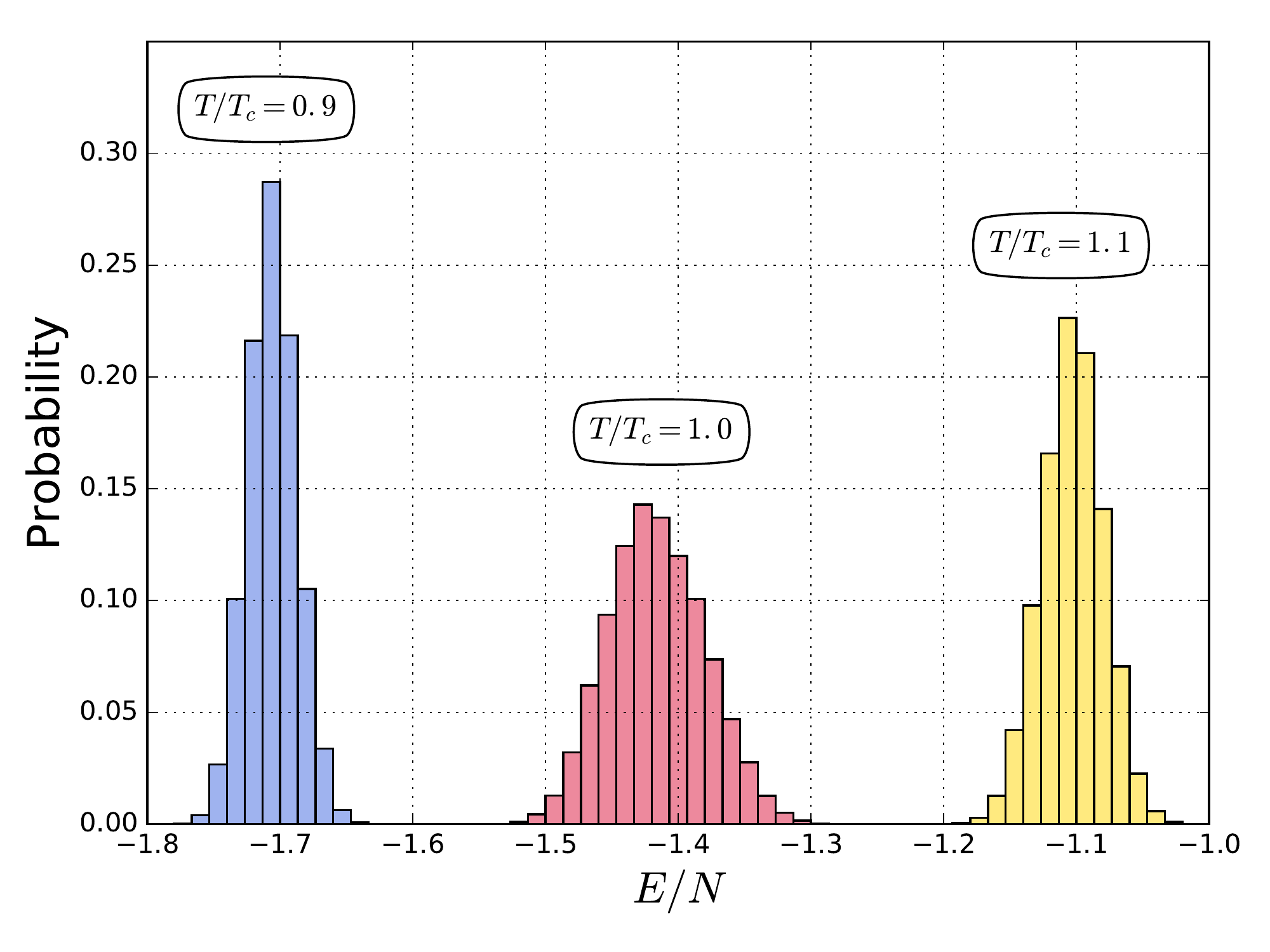}

\caption{Histograms of the energy per ion site obtained at different temperatures
in a ferromagnetic 2D Ising system.\label{fig:histograms}}

\end{figure}

From results obtained using \noun{Vegas,} it is also possible to apply
data analysis techniques such as the single and multiple histogram
method. These methods allow to take a single simulation performed
at some temperature and extrapolate or interpolate results to give
predictions of observable quantities at other temperatures \citep{Newman1999}.
Figure \ref{fig:histograms} show histograms of the energy per ion
site ($E/N$) obtained at different temperatures in a ferromagnetic
2D Ising system with generic parameters. Using the multiple histogram
method, it is possible to interpolate observable quantities, such
as the energy and magnetization, at other temperatures with high accuracy.

\subsection{Identification of magnetization states}

\begin{figure}[h]
\begin{centering}
\subfloat[]{\includegraphics[width=0.3\columnwidth]{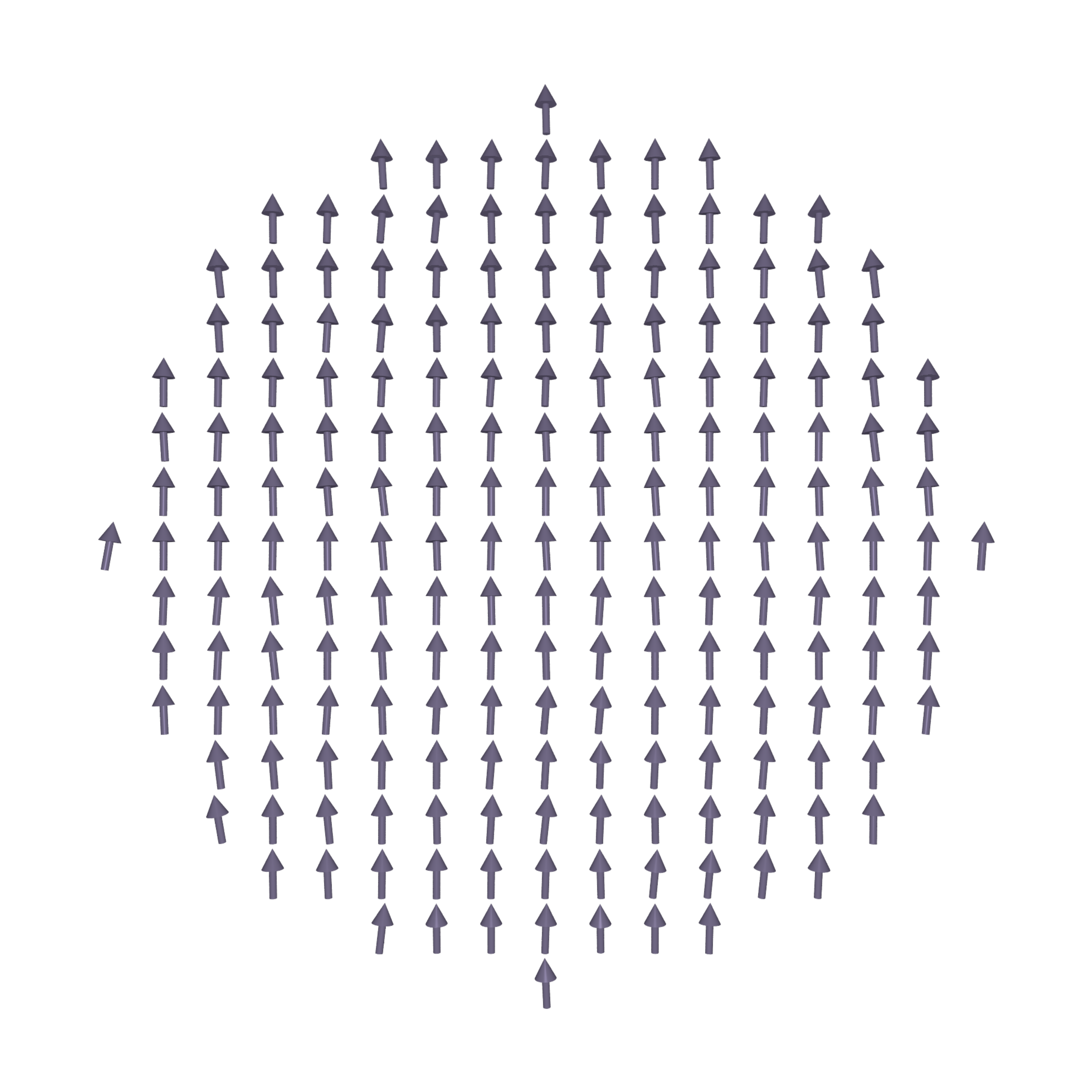}

}\subfloat[]{\includegraphics[width=0.3\columnwidth]{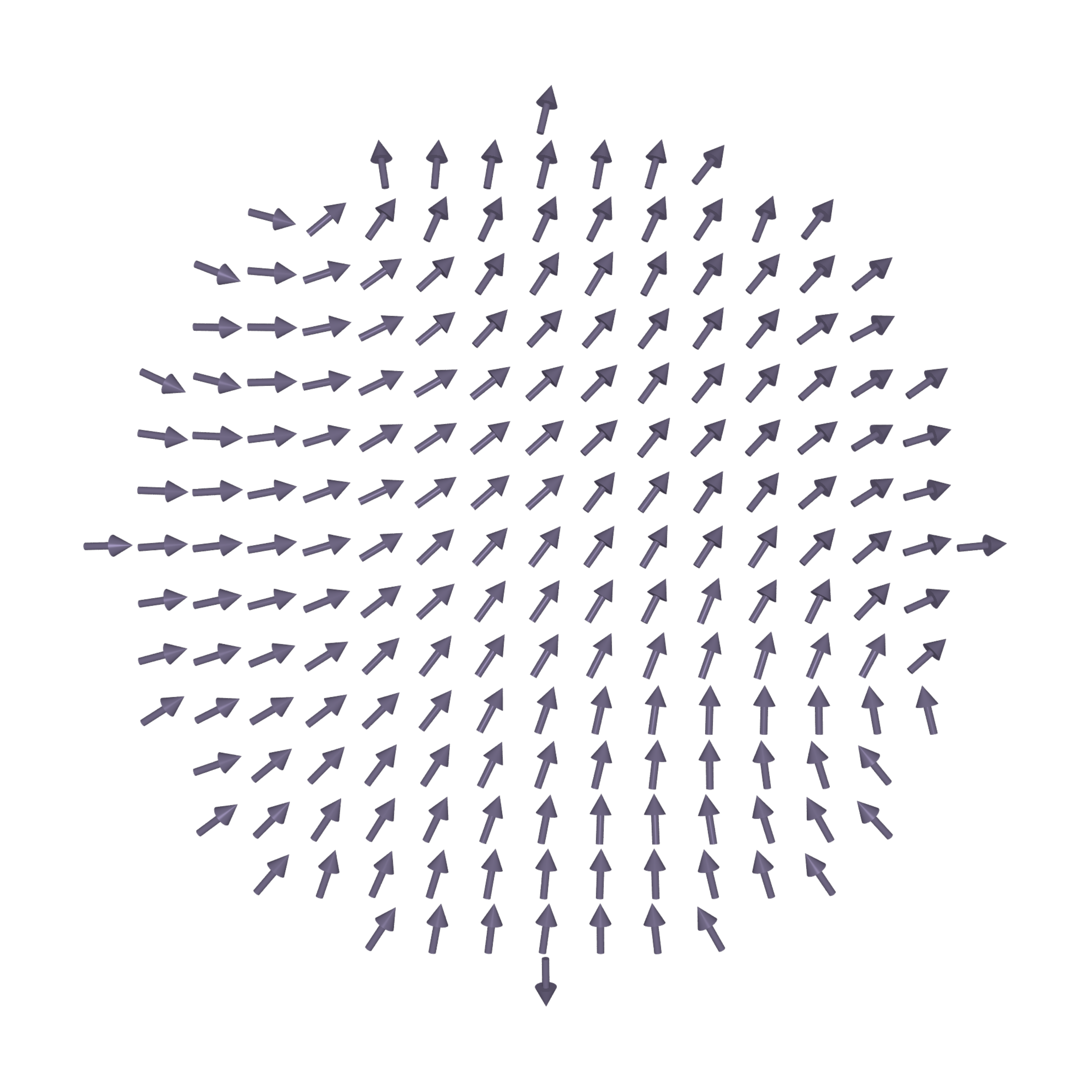}

}
\par\end{centering}
\begin{centering}
\subfloat[]{\includegraphics[width=0.3\columnwidth]{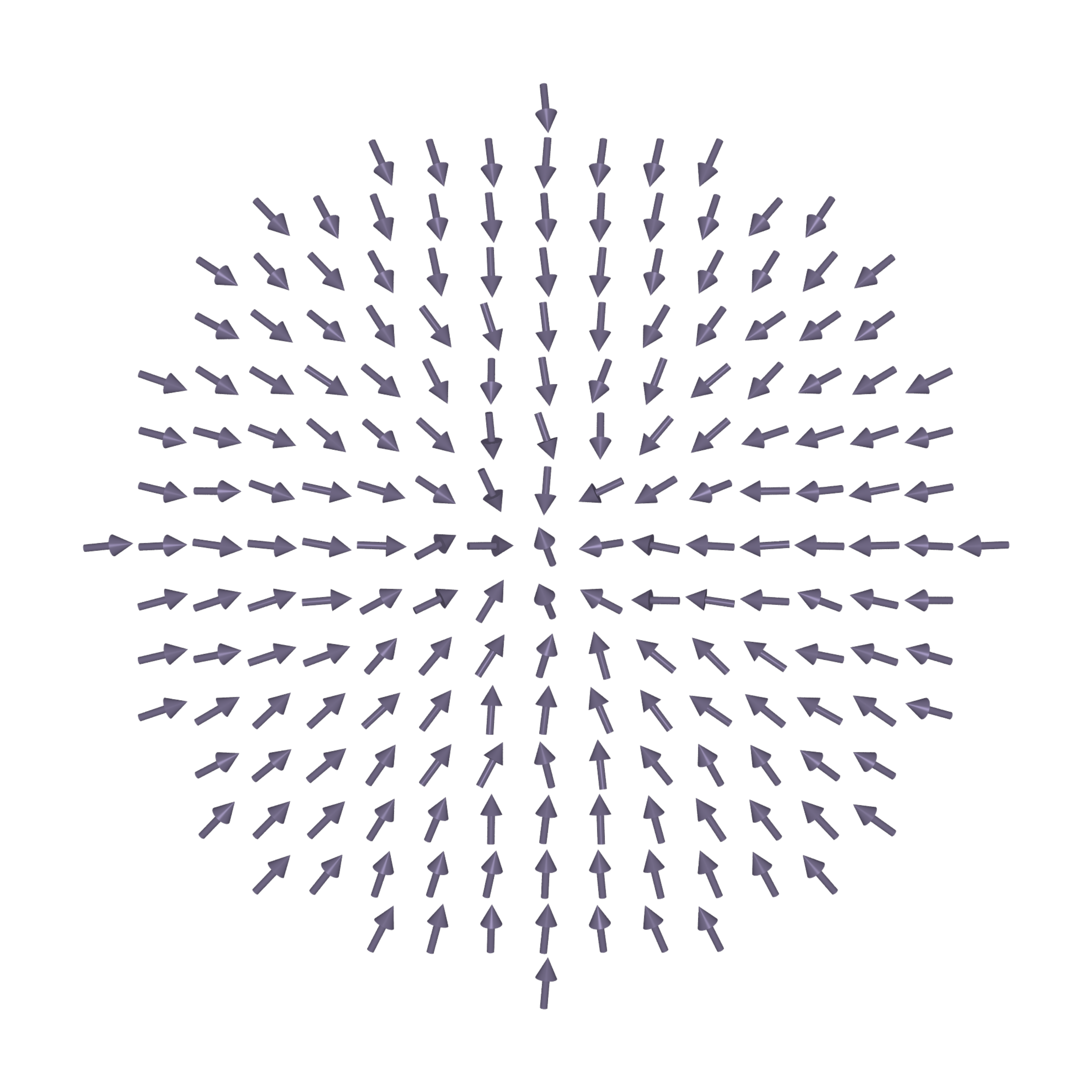}

}\subfloat[]{\includegraphics[width=0.3\columnwidth]{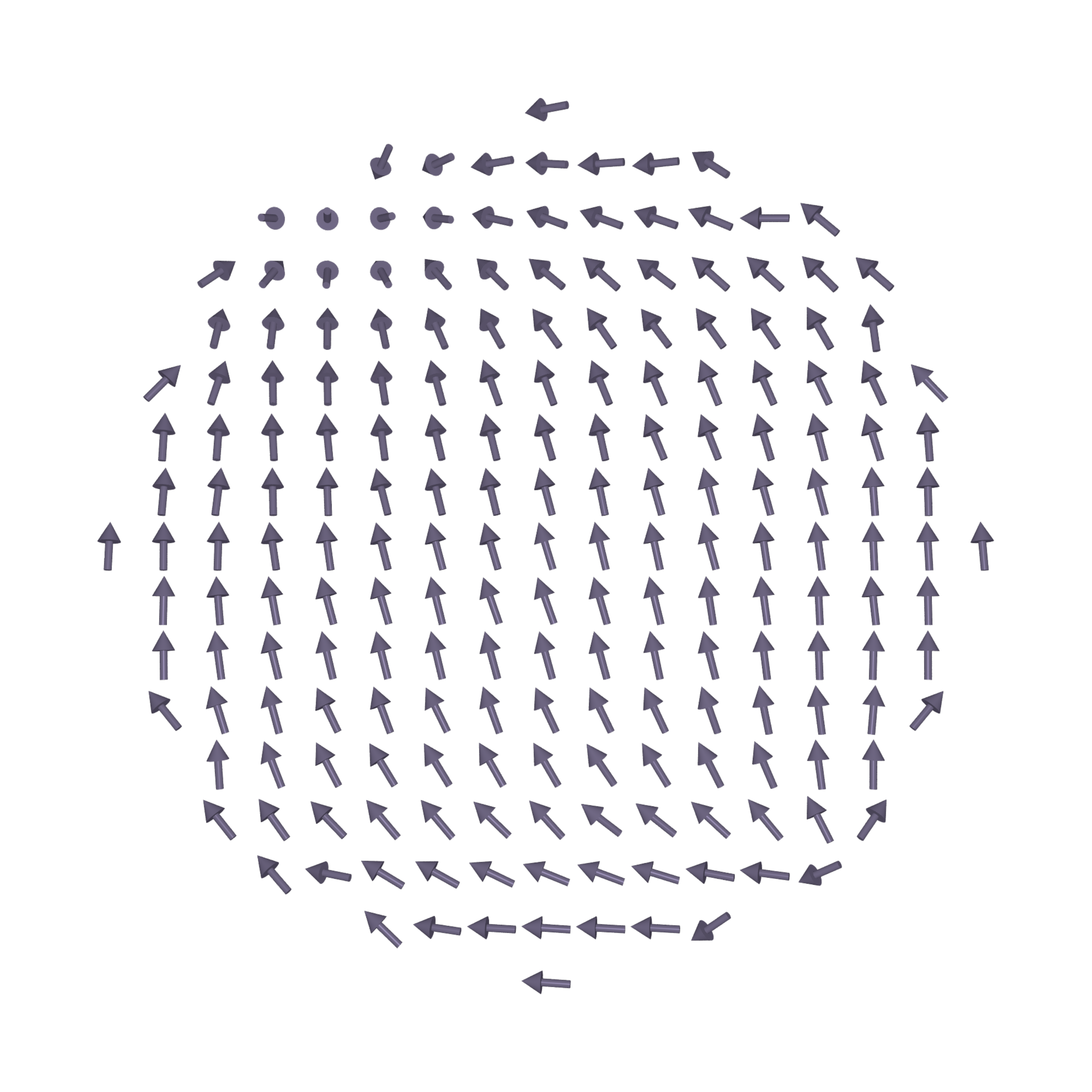}

}
\par\end{centering}
\caption{(a) ``Collinear'', (b) ``throttled'', (c) ``hedgehog'' and (d)
``artichoke'' magnetization states produced in a ferromagnetic spherical
nanoparticle with core and Néel surface anisotropy. \label{fig:states_nanoparticle}}
\end{figure}

The feature of \noun{Vegas} that allows to store the direction of
every spin moment at the last MCS is very useful to identify different
magnetization states, which is difficult from the thermal average
quantities alone. Figure \ref{fig:states_nanoparticle} shows a cross-sectional
view of different magnetization states produced in a spherical ferromagnetic
nanoparticle with core ($k_{c}$) and Néel surface ($k_{s}$) anisotropy.
The magnetization state depends on the ratio $k_{s}/k_{c}$. When
$k_{s}/k_{c}\rightarrow0$, $100$, $\infty$ and $-100$, ``collinear'',
``throttled'', ``hedgehog'' and ``artichoke'' states are produced,
respectively \citep{Alzate-Cardona2017b}.

Figures \ref{fig:materials}, \ref{fig:torus_a}, \ref{fig:torus_b},
\ref{fig:states_nanoparticle} and the inset in Figure \ref{fig:irona}
were generated using \noun{POV-Ray} \citep{Ltd.2004}. 

\section{Perspectives}

\noun{Vegas} is a software package under active development. Important
improvements are planned for the sample building and data analysis
libraries, increasing the flexibility in the construction of complex
magnetic systems and providing different tools for the analysis and
visualization of all kind of simulation results. Addition of new Hamiltonian
terms, such as the dipolar and Dzyaloshinskii-Moriya interactions,
are expected to be included in the Hamiltonian. Parallelization techniques
are also planned to be applied to the Monte Carlo Metropolis algorithm
to reduce simulation times. Researchers are encouraged to contribute
to the development of \noun{Vegas} at its GitHub repository \citep{Alzate-Cardona}.

\section{Acknowledgments}

Simulations in this work were done using resources provided by the
Open Science Grid \citep{Pordes2007}, which is supported by the National
Science Foundation and the U.S. Department of Energy's Office of Science.
We acknowledge support by COLCIENCIAS under the program Jóvenes Investigadores
e Innovadores 2016 (Grant No. 761).

\bibliographystyle{ieeetr}
\bibliography{manuscript}

\end{document}